\documentclass[aps,12pt]{revtex4}
\usepackage[sort&compress]{natbib}
\usepackage{xcolor}
\usepackage{graphicx}
\usepackage{amsmath}
\usepackage{amssymb}
\usepackage{epstopdf}
\usepackage{caption}
\usepackage{float}
\usepackage{hyperref}
\hypersetup{
    colorlinks,
    citecolor=blue,
    filecolor=blue,
    linkcolor=blue,
    urlcolor=blue}    
\begin{document}

\title{Impact of temperature asymmetry and small fraction of static positive ions on the relaxed states of a relativistic hot pair plasma}

\author{Usman Shazad}
\email{usmangondle@gmail.com}
\author{M. Iqbal}
\affiliation{Department of Physics, University of Engineering and Technology, Lahore
54890, Pakistan}

\begin{abstract}
The relaxed state of a magnetized relativistic hot plasma composed of inertial electrons and positrons
having different relativistic temperatures and a fraction of static positive ions is studied. From the steady-
state solutions of vortex dynamics equations and the relation for current density, a non-force-free triple
Beltrami (TB) relaxed state equation is derived. The TB state is characterized by three scale parameters that
consequently provide three different self-organized structures. The analysis of the relaxed state shows that
for specific values of generalized helicities, the disparity in relativistic temperature and the existence of a
small fraction of static positive ions in pair plasma can transform the nature of scale parameters. Moreover,
an analytical solution of the TB state for an axisymmetric cylindrical geometry with an internal conductor
configuration demonstrates that due to asymmetries of temperature and density of plasma species,
diamagnetic structures can transform into paramagnetic ones and vice versa. The present study will improve our understanding of pair plasmas in trap-based plasma confinement experiments and astrophysical environments.
\end{abstract}
\maketitle
\section{Introduction}\label{S1}

Plasmas consisting of relativistic electron-positron (EP) pairs are prevalent in various high-energy astrophysical phenomena, including the early universe as well as the vicinity of neutron stars and black holes \cite{Holcomb1989,Ginzburg1971,Sturrock1971,Arons1979,Blandford1995,Wardle1998,Mirabel1999,Meszaros2002,Weidenspointner2008}. The EP plasmas in these astrophysical environments exhibit relativistic temperatures and can also attain relativistic velocities \cite{Blandford1977}. In the context of relativistic hot plasma, the thermal energy possessed by the plasma particles is either equivalent to or surpasses their rest mass energy. Important examples believed to be associated with relativistic hot 
pair plasmas include pulsar magnetospheres \cite{Spitkovsky2006,Brambilla2018}, gamma-ray bursts (GRB) \cite{Geng2016}, and outflows from active galactic nuclei (AGN) \cite{Bottcher1997,Blandford2019}. It is also believed that in these relativistic EP plasmas, a small fraction of positive ions can also exist, thus forming an electron-positron-ion (EPI) plasma \cite{Berezhiani1994}.
In contemporary times, it is possible to create EP or EPI plasmas within laboratory settings. This can be achieved through various means, such as the direct exposure of high-atomic-number metal targets to intense laser pulses or the utilization of a relativistic electron beam produced by accelerating electrons in a laser wakefield \cite{Sarri2015}.
In recent years, extensive research has been conducted on relativistic hot plasmas in the context of astrophysical phenomena and laser-plasma interactions, owing to their potential significance. As a result, several non-linear plasma phenomena and structures have been thoroughly investigated \cite{Berezhiani1995,Pokhotelov2001,Lopez2012,Asenjo2012,Comisso2014,Mahajan2018}.

In addition to the presence of non-linear structures and processes, such as waves, solitons, and instabilities \cite{Saha2021,Chatterjee2023}, the phenomenon of magnetic self-organization (MSO), commonly referred to as relaxation, holds significant importance in both natural and laboratory plasmas. Historically, the process of MSO was dealt with employing the variational principle, wherein the minimization of ideal invariants of the plasma system under certain constraints results in a self-organized or relaxed state \cite{Ortolani}. For instance, in ideal magnetohydrodynamics (MHD), when the magnetic 
energy of plasma under magnetic helicity constraint is minimized, the relaxed 
state can be expressed as $\mathbf{\nabla }\times \mathbf{B}=\lambda \mathbf{B}$, 
where $\mathbf{B}$ is the magnetic field and $\lambda$ is a constant and eigenvalue 
of the curl operator and also called scale parameter. Moreover, $\lambda$ is a measure of shear and twist in the magnetic field ($\because$ $\mathbf{\nabla }\times \mathbf{B}\cdot \mathbf{B}/B^{2}$) and the size of the 
relaxed state structure because its dimensions are equal to the inverse of length. It is also noteworthy to mention that a vector field exhibiting the alignment of its vortex with itself is referred to as a Beltrami field. That is why this relaxed state is commonly referred to as the Beltrami state. The single Beltrami relaxed state of an ideal MHD with constant $\lambda$ is a
linear force-free state that is flowless and devoid of pressure gradients \cite{Woltjer1958,Taylor1974}. 
Contrary to this, when $\lambda(r)$ is some scalar function, it represents a non-linear force-free state.

Later on, the more realistic, non-force-free relaxed states, which also show significant pressure gradients and magneto-fluid coupling, are modeled in Hall MHD  (HMHD) through the utilization of vortex dynamics and variational approaches. The relaxed state of a HMHD plasma is the double Beltrami (DB) state, which can be expressed as the linear sum of two single force-free states and is characterized by two scale parameters \cite{Mahajan1998,Steinhauer2002}. On the other hand, for two-fluid plasmas, when the inertial effects of both plasma species are considered, the relaxed state is the triple Beltrami (TB) state, which is the linear superposition of three single Beltrami states, and three scale parameters are associated with this relaxed state \cite{Bhattacharyya2003,Iqbal2013APL}. More recently, it has been shown that by taking into account the inertial effects of all the plasma species in relativistic degenerate three-component EPI plasma, the relaxed state is a quadruple Beltrami (QB) state, which is a combination of four single force-free fields and characterized by four self-organized vortices \cite{Shatashvili2016}.

In recent years, there has been growing interest among researchers in the field of plasma relaxation, and several researchers have explored the Beltrami equilibrium states in different astrophysical plasma environments, including dense degenerate \cite{Berezhiani2015,Shatashvili2016,Shatashvili2019} and classical relativistic hot plasmas \cite{Iqbal2008,Pino2010,Iqbal2013,Usman2021,Usman2023}, plasmas surrounding black holes with the inclusion of general relativistic effects \cite{Asenjo2019,Bhattacharjee2020}, planetary multi-ion plasmas \cite{Shafa2022}, and partially ionized dusty plasmas \cite{Faheem2023}. In a recent study, Bhattacharjee has expanded the notion of Beltrami equilibrium states to self-gravitating neutral fluids and magnetized plasmas in massive photon fields. For instance, the equilibrium state of a weakly rotating self-gravitating neutral fluid while being subjected to a stationary gravitomagnetic field is a DB state, while the relaxed state of a single fluid magnetized plasma by incorporating the photon mass is a TB state
\cite{Bhattacharjee2022,Bhattacharjee2023}.
Importantly, the above-mentioned multi-Beltrami relaxed states have been extensively used to model various laboratory and astrophysical phenomena, such as 
high beta relaxed states in tokamak \cite{Yoshida2001}, field reverse configurations \cite{Bhattacharyya2001}, heating of solar corona \cite{Mahajan2001}, eruptive events in
solar plasma \cite{Ohsaki2002,Kagan2010}, solar arcades \cite{Bhattacharyya2007}, and dynamo mechanisms 
\cite{Mahajan2005,Kotorashvili2020,Kotorashvili2022}.

The goal of the present study is to investigate the TB relaxed states of two-temperature relativistic hot EP plasma with a small fraction of immobile positively charged ions. So there are two asymmetries between pair species in this plasma system: one is the different effective inertia's (due to different relativistic temperatures) for the electron and positron species, and the other is the presence of a small number of static ions. Also, an analytical solution of the TB state in axisymmetric cylindrical geometry with an internal conductor configuration (annular plasma system) is presented. Then, it is demonstrated that the asymmetries in density and thermal energy of plasma species can transform diamagnetic structures into paramagnetic ones. The originality of the present work resides in the asymmetry between the relativistic temperatures of electron and positron species. In previous studies, it was presumed that the relativistic temperatures of both plasma species were equal \cite{Iqbal2008,Pino2010,Iqbal2013,Usman2021,Usman2023}. Another novelty is that both paramagnetic and diamagnetic structures can be created in the TB state with an internal conductor configuration. However, previous investigations of TB states in electron-ion and multi-ion plasma have demonstrated that only paramagnetic structures are possible \cite{Gondal2021,Gondal2022}.

It is also important to mention that the generation of relativistic pair plasmas in laboratory settings is increasingly attracting the attention of scientists. In a recent study conducted by Chen et al. at Lawrence Livermore National Laboratory, positrons were generated with temperatures distinct from those of electrons \cite{Chen2009,Chen2010,Chen2011}. Moreover, a number of recent studies have been conducted by Mahajan et al. and Berezhiani et al., focusing on the examination of relativistic plasma featuring temperature asymmetry. The aforementioned studies also proposed the possibility of a temperature asymmetry emerging when the plasma species are generated under different conditions \cite{Mahajan2009,Berezhiani2010,Berezhiani2010a,Berezhiani2013}. In laboratories, the creation of two-temperature relativistic pair plasmas has become feasible as a result of recent advancements in lasers, trap design, confinement, positron accumulation, and injection \cite{Sarri2015,Stoneking2020,Fajans2020,Linden2021,Chen2023}.

Furthermore, it is crucial to highlight that the internal conductor or annular plasma configurations are basically trap-based configurations, which are the subject of intensive research. For example, the toroidal trap device, namely Ring Trap-1 (RT-1) at the University of Tokyo, also called the laboratory magnetosphere, is equipped with a superconducting internal coil. The mechanism for plasma confinement in the RT-1 plasma device is based on the concept of MSO and recently Beltrami-Bernoulli states are investigated in this device \cite{Yoshida2006,Yoshida2010b}. Most recently, the steady-state of EP plasma in the RT-1 has also been investigated \cite{Sato2023}. Some other examples of such plasma configurations are the magnetic reconnection experiment (MRX) facility at Princeton plasma physics laboratory (PPPL) and galatea traps \cite{Yamada1997,Morozov2007}.

The paper is organized in the following manner: In Sec. \ref{S2}, from
macroscopic evolution equations, the TB equation is derived. The characteristics
of scale parameters are described in Sec. \ref{S3}. In Sec. \ref{S4}, the analytical solution of the TB state in cylindrical geometry with an internal conductor is presented and the impact of temperature and species density asymmetries on the relaxed state structures are explored. The summary of the present study is provided in Sec. \ref{S5}.

\section{Model equations and TB state}\label{S2}

We consider a magnetized, collisionless and incompressible three-component
plasma which is composed of electrons, positrons and a fraction of
positively charged ions.The electrons and positrons are dynamic and relativistic, with different temperatures, while the ions are static and play their role only in quasineutrality and charge separation.  Now, the
quasineutrality condition for the plasma system can be stated as 
\begin{equation}
N_{p}+N_{i}=1,
\end{equation}
where $N_{p}=n_{0p}/n_{0e}$ and $N_{i}=z_{i}n_{0i}/n_{0e}$, in which $n_{0e}$, $n_{0p}$ and  $n_{0i}$ are the electron, positron and ion number densities in the rest frame whereas $z_{i}$ is charge state of ion species. If the velocity distribution of plasma species is local relativistic
Maxwellian, then by following the Ref. \cite{Berezhiani2002}, the equation
of motion for relativistic plasma species can be expressed as
\begin{equation}
\frac{\partial \mathbf{\Pi}_{\alpha }}{\partial t}+\frac{1}{n_{\alpha }}
\mathbf{\nabla }\mathit{p}_{\alpha }=q_{\alpha }\mathbf{E}+\mathbf{V}
_{\alpha }\times \left(\mathbf{\Pi}_{\alpha }+\frac{q_{\alpha }}{c}\mathbf{B}
\right),  \label{me}
\end{equation}
where $\alpha $ stands for electrons ($e$) and positrons ($p$) while $%
\mathbf{\Pi}_{\alpha }=\gamma _{\alpha }m_{0\alpha }G_{\alpha }\mathbf{V}%
_{\alpha }$, $\mathbf{V}_{\alpha }$, $m_{0\alpha }$, $q_{\alpha }$, $c$, $\mathit{p}_{\alpha}=\left( n_{\alpha }/\gamma _{\alpha }\right) T_{\alpha }$, $%
n_{\alpha }$, $T_{\alpha }$, $G_{\alpha }$ and $\gamma _{\alpha }=\left(
1-V_{\alpha }^{2}/c^{2}\right) ^{-1/2}$ are relativistic momentum, velocity,
rest mass, electric charge, speed of light, relativistic pressure, number density, proper
temperature, relativistic temperature factor and relativistic Lorentz factor, respectively. Moreover, in above equation of motion, $%
\mathbf{E}=\mathbf{-\nabla }\phi -c^{-1}\partial \mathbf{A/\partial }t$ and $%
\mathbf{B}=\mathbf{\nabla }\times \mathbf{A}$ are electric and magnetic
fields that are related to scalar and vector potentials $\phi $ and $\mathbf{%
A}$, respectively. As mentioned above that a plasma is said to be relativistic when the directed
fluid velocity approaches the speed of light or the thermal energy of plasma
particles is equal or larger than the rest mass energy. Both approaches to
relativity emerge in astrophysical and laboratory plasmas. However, the present study solely
accounts for thermal relativistic effects, wherein the thermal energy of
plasma particles is either equivalent to or greater than their rest mass
energy, while the fluid velocity of plasma species is non-relativistic ($%
\gamma _{\alpha }\approx 1$). As mentioned above the velocity distribution
of plasma species is local relativistic Maxwellian, then the thermal
relativistic effects appear through the factor $G_{\alpha }(z_{\alpha
})=K_{3}(z_{\alpha }^{-1})/K_{2}(z_{\alpha }^{-1})$, where $K_{2}$ and $K_{3}
$ are the modified Bessel functions and $z_{\alpha }=T_{\alpha }/m_{0\alpha
}c^{2}$. For non-relativistic case $z_{\alpha }<<1$, and the factor $%
G(z_{\alpha })$ can approximately be taken as $G(z_{\alpha })\approx
1+5z_{\alpha }/2$, while on the other hand, for highly relativistic plasma
species $z_{\alpha }>>1$, and the relativistic temperature $G(z_{\alpha })$
can be approximated as $G(z_{\alpha })\approx 4z_{\alpha }$. At this point, it is also very important to make it clear that our plasma system is stable. This is because the effective collision frequency in the relativistic hot EP plasma, which includes recombination and photon annihilation, is thought to be much lower than the plasma frequency. In a study by Tajima and Taniuti have extensively examine the conditions under which wave equations and collective plasma processes remain valid in a similar physical setting \cite{Tajima1990}
 
Now the equations of motion for the relativistic hot two-temperature
electrons and positrons in normalized form can be expressed as 
\begin{equation}
\frac{\partial }{\partial t}\left( G_{e}\mathbf{V}_{e}-\mathbf{A}%
\right) =\mathbf{V}_{e}\times \left( \mathbf{\nabla }\times G_{e}%
\mathbf{V}_{e}-\mathbf{B}\right) -\mathbf{\nabla }\psi _{e},
\label{e21}
\end{equation}
\begin{equation}
\frac{\partial }{\partial t}\left( G_{p}\mathbf{V}_{p}+\mathbf{A}%
\right) =\mathbf{V}_{p}\times \left( \mathbf{\nabla }\times G_{p}%
\mathbf{V}_{p}+\mathbf{B}\right) -\mathbf{\nabla }\psi _{p},
\label{e22}
\end{equation}
where $G_{e}$, $G_{p}$, $\mathbf{V}_{e}$ and $\mathbf{V}_{p}$ are
relativistic temperatures and velocities of electron and positron species,
respectively. For the normalization of length, time, magnetic
field, velocity and temperature of plasma species in equations (\ref{e21}-\ref{e22}), we used
the following plasma parameters: electron skin depth $%
\lambda _{e}=\sqrt{m_{0}c^{2}/4\pi e^{2}n_{0e}}$ (where $m_{0}$ is rest mass of electron),  $
t=\lambda _{e}/v_{A}$, some arbitrary magnetic field $%
B_{0}$, Alfv\'{e}n speed $v_{A}=B_{0}/\sqrt{4\pi m_{0}n_{0e}}$ and rest mass
energy of electron $m_{0}c^{2}$, respectively.
Whereas, the gradient terms in above equations have the following values: $%
\psi _{e,p}=-c_{a}^{2}G_{e,p}\mp \phi $, in which $c_{a}=c/v_{A}$.

Now, in order to derive the vorticity evolution equations, we take the curl
of the equations (\ref{e21}-\ref{e22}). The gradient terms are eliminated ($\mathbf{\nabla }\times\mathbf{\nabla }\psi _{e,p}=0$), and we are
left with the vortex dynamics equations for the electron and positron
species as given below 
\begin{equation}
\frac{\partial \mathbf{\Omega }_{\alpha}}{\partial t}=\mathbf{\nabla }\times
\left( \mathbf{V}_{\alpha}\times \mathbf{\Omega }_{\alpha}\right) ,  \label{e3}
\end{equation}
where $\mathbf{\Omega }_{\alpha=e,p}=\nabla \times G_{\alpha}\mathbf{V}_{\alpha}\mp 
\mathbf{B}$ is called generalized or canonical vorticity of plasma species. One can obtain two
Beltrami conditions from the steady-state solution of the vorticity
evolution equations ($\mathbf{V}_{\alpha}\times \mathbf{\Omega }_{\alpha }=0$),
which assert that the generalized vorticity is parallel to its associated
flow ($\mathbf{\Omega }_{\alpha}\parallel \mathbf{V}_{\alpha}$). So from equation (\ref{e3}), the Beltrami conditions are given by 
\begin{equation}
\nabla \times G_{e}\mathbf{V}_{e}-\mathbf{B}=aG_{e}%
\mathbf{V}_{e},  \label{e41}
\end{equation}\begin{equation}
\nabla \times G_{p}\mathbf{V}_{p}+\mathbf{B}=bG_{p}%
\mathbf{V}_{p},  \label{e42}
\end{equation}
where $a$ and $b$ are called Beltrami parameters and are the ratios of generalized vorticities to corresponding flows, as well as the measure of generalized helicities of plasma species. The generalized helicities are ideal invariants of the plasma system, which are discussed at the end of this section. Moreover, to close the system of equations for the derivation of the relaxed state equation, we use Ampere's law. So the
Ampere's law in normalized form can be expressed as 
\begin{equation}
\mathbf{\nabla }\times \mathbf{B}=N_{p}\mathbf{V}_{p}-\mathbf{V}_{e}.
\label{e5}
\end{equation}
Now, in order to represent the Beltrami conditions in terms of
composite flow velocity ($\mathbf{V}$) and magnetic field, we express
electron and positron velocities in terms of $\mathbf{V}$. The relation for
composite flow velocity can be written as 
\begin{equation}
\mathbf{V}=\xi ^{-1}\left( \mathbf{V}_{e}+N_{p}\mathbf{V}_{p}\right) ,
\label{e6}
\end{equation}
where $\xi=1+N_{p}$. From equations (\ref{e5}-\ref{e6}), the velocities of
the electron and positron species are given by 
\begin{equation}
\mathbf{V}_{e}=\frac{\xi \mathbf{V-\mathbf{\nabla }\times \mathbf{B}}}{2},
\label{e7}
\end{equation}
\begin{equation}
\mathbf{V}_{p}=\frac{\xi \mathbf{V}+\mathbf{\nabla }\times \mathbf{B}}{2N_{p}
}.  \label{e8}
\end{equation}
Using equations (\ref{e7}-\ref{e8}) in Beltrami conditions (\ref{e41}-\ref{e42}),
one can obtain following two equations in terms of $\mathbf{B}$ and $\mathbf{V}$
\begin{equation}
\xi \left( b\mathbf{V}-\mathbf{\nabla }\times \mathbf{V}\right) -(
\mathbf{\nabla }\times )^{2}\mathbf{B}+b\mathbf{\nabla }\times \mathbf{B}%
=2N_{p}G_{p}^{-1}\mathbf{B},  \label{e9}
\end{equation}
\begin{equation}
\xi \left( \mathbf{\nabla }\times \mathbf{V}-a\mathbf{V}\right) -(
\mathbf{\nabla }\times )^{2}\mathbf{B}+a\mathbf{\nabla }\times \mathbf{B}
=2G_{e}^{-1}\mathbf{B},  \label{e10}
\end{equation}
where $\mathbf{\nabla }\times \mathbf{\nabla }\times =(\mathbf{\nabla }%
\times )^{2}$. Solving above equations, the composite velocity $\mathbf{V}$
comes out to be
\begin{equation}
\mathbf{V}=f_{3}(\mathbf{\nabla }\times )^{2}\mathbf{B}-f_{2}\mathbf{\mathbf{
\mathbf{\nabla }\times \mathbf{B}}}+f_{1}\mathbf{\mathbf{\mathbf{B,}}}
\label{e11}
\end{equation}
where $f_{3}=2\kappa $, $f_{2}=\kappa \left( a+b\right) $, $%
f_{1}=2\kappa \left( G_{e}^{-1}+N_{p}G_{p}^{-1}\right) $ and $\kappa
=1/(b-a)\xi $. By eliminating $\mathbf{V}$ from equations (\ref{e9}-\ref{e10} ), we obtain
the following relaxed state equation for magnetic field 
\begin{equation}
(\mathbf{\nabla }\times )^{3}\mathbf{B}-k_{3}(\mathbf{\nabla }\times )^{2}%
\mathbf{B}+k_{2}\mathbf{\mathbf{\mathbf{\mathbf{\nabla }\times B}}}-k_{1}%
\mathbf{B}=0,  \label{TB}
\end{equation}
where $\mathbf{\nabla }\times \mathbf{\nabla }\times \mathbf{\nabla }\times
=(\mathbf{\nabla }\times )^{3}$, $k_{3}=a+b$,$\
k_{2}=ab+G_{e}^{-1}+N_{p}G_{p}^{-1}$ and $%
k_{1}=bG_{e}^{-1}+aN_{p}G_{p}^{-1}$. The equation (\ref{TB}) describing the relaxed state of the relativistic hot two-temperature EPI plasma is a TB equation and in the next section we will show that it can be expressed as a superposition of three distinct Beltrami fields. The equations (\ref{e11}-\ref{TB}) also show a strong coupling
between flow and field in the relaxed state.  Importantly, if the relativistic temperatures of both species are equal in the absence of ion species, previous results from Ref. \cite{Iqbal2008} can be derived from the relaxed state equation (\ref{TB}). However, for same temperatures with static positive ion species, the results of Ref. \cite{Usman2021} can be obtained. Consequently, the present study is more comprehensive and accounts for both temperature and density asymmetry.  Also noteworthy is the fact that relaxed state equation (\ref{TB}) can also be derived using the variational technique. Three constants of motion, namely the generalized helicities of pair species ($h_{e}$ and $h_{p}$) and magnetofluid energy ($E_{mf}$), can be derived from equations (\ref{e21}-\ref{e3} and \ref{e5}) for this plasma system that can be expressed as
\begin{equation}
    h_{e}=0.5\left\langle \text{curl}^{-1}\mathbf{\Omega}_{e
}\cdot \mathbf{\Omega }_{e}\right\rangle,
\end{equation}
\begin{equation}
    h_{p}=0.5\left\langle \text{curl}^{-1}\mathbf{\Omega}_{p
}\cdot \mathbf{\Omega }_{p}\right\rangle,
\end{equation}
\begin{equation}
    E_{mf}=0.5\left\langle
G_{e}V_{e}^{2}+N_{p}G_{p}V_{p}^{2}+B^{2}\right\rangle, \label{energy}
\end{equation}
where $\left\langle
-\right\rangle $ represents volume integral \cite{Steinhauer1997}. Moreover, it can be observed that these ideal invariants provide confirmation for the presence of n+1 ideal invariants in the case of n inertial species, as well as the assertion that the relaxed state is a TB state \cite{Mahajan2015}.

\section{Characteristics of scale parameters}\label{S3}

The commutative nature of the curl operators enables us to express the TB equation (\ref{TB}) as the linear superposition of three distinct Beltrami fields $\mathbf{B}%
_{j}$ ($j=1,2,3$). These Beltrami fields fulfill the
following condition: $\mathbf{\nabla }\times \mathbf{B}_{j}=\lambda _{j}%
\mathbf{B}_{j}$, where $\lambda _{j}$ and $\mathbf{B}_{j}$ are eigenvalues
and eigenfunctions of the curl operator, respectively \cite{Yoshida1990}. Now, by introducing the concept of eigenvalues, it is possible to express the equation (\ref{TB}) in the
following manner: 
\begin{equation}
(\text{curl}-\lambda _{_{1}})(\text{curl}-\lambda _{2})(\text{curl}-\lambda
_{3})\mathbf{B}=0,  \label{e12}
\end{equation}
where $\lambda _{1}$, $\lambda _{2}$ and $\lambda _{3}$ are the eigenvalues that are also called scale parameters and are associated with three distinct Beltrami fields. The coefficients ($k_{j}$) of TB equation (\ref{TB}) and eigenvalues ($\lambda _{j}$) from equation (\ref{e12}) can be related as 
\begin{eqnarray*}
k_{1} &=&\lambda _{1}\lambda _{2}\lambda _{3}, \\
k_{2} &=&\lambda _{1}\lambda _{2}+\lambda _{1}\lambda _{3}+\lambda
_{2}\lambda _{3}, \\
k_{3} &=&\lambda _{1}+\lambda _{2}+\lambda _{3}.
\end{eqnarray*}
The relationships shown above between the coefficients of the TB equation (\ref{TB}) and the eigenvalues demonstrate that these are the roots of the
following cubic equation 
\begin{equation}
\lambda ^{3}-k_{3}\lambda ^{2}+k_{2}\lambda -k_{1}=0.  \label{e13}
\end{equation}
The roots of equation (\ref{e13}) are as follows 
\begin{equation}
\lambda _{1}=\frac{a+b}{3}+P+Q,
\end{equation}
\begin{equation}
\lambda _{2}=\frac{a+b}{3}-\frac{1}{2}\left( P+Q\right) +\frac{i
\sqrt{3}}{2}\left( P-Q\right) ,
\end{equation}
\begin{equation}
\lambda _{3}=\frac{a+b}{3}-\frac{1}{2}\left( P+Q\right) -\frac{i
\sqrt{3}}{2}\left( P-Q\right),
\end{equation}
where $P=\left( \left( S/2\right) -\sqrt{D}\right) ^{1/3}$, $Q=\left( \left(
S/2\right) +\sqrt{D}\right) ^{1/3}$, $D=\left( S^{2}/4\right) +\left(
T^{3}/27\right) $, $S=\left( u_{1}u_{2}/27\right) -\left(
u_{1}/3G_{e}\right) +\left( u_{3}N_{p}/3G_{p}\right) $, $%
T=N_{p}G_{p}^{-1}+G_{e}^{-1}+\left( u_{4}/3\right) $, $u_{1}=a+2b$, $%
u_{2}=2a^{2}+ab-b^{2}$, $u_{3}=2a-b$, $%
u_{1}=a+2b$, and $u_{4}=ab-a^{2}-b^{2}$. 
\begin{figure*}
    \centering
    \includegraphics[scale=0.9]{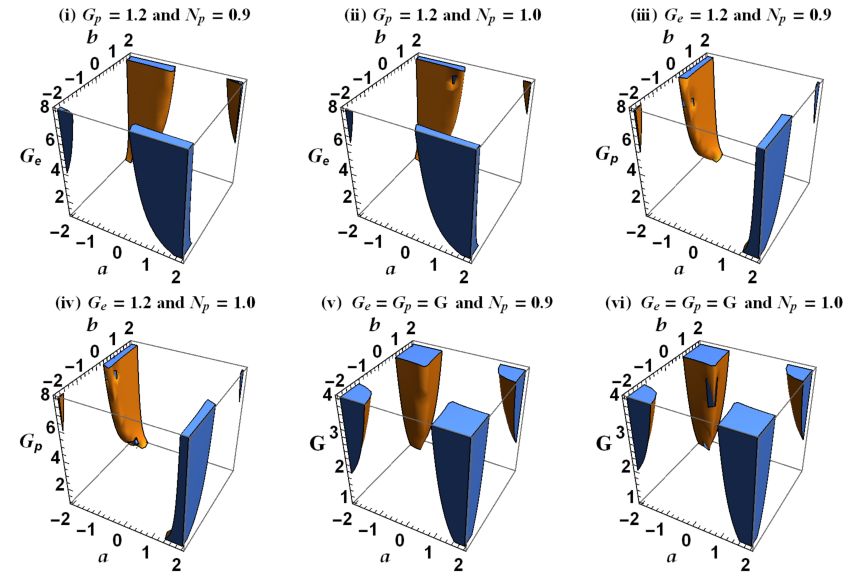}
    \caption{Plot of $\Delta>0$, as a function of Beltrami parameters ($a$ and $b$) and relativistic temperature of pair species ($G_{e}$ and $G_{p}$) for different plasma settings. In the colored region, all the scale parameters are distinct and real-valued, while in the transparent region, one scale parameter is real and the other two are complex conjugate pairs.}
    \label{fig:1}
\end{figure*}

As previously stated, the scale parameters have dimensions that are the reciprocal of length. Therefore, these scale parameters play a crucial role in determining the size of the relaxed state structures. Additionally, it is worth noting that the scale parameters can take on values that are either real, complex, or imaginary. Consequently, the character of the relaxed state depends on the nature of the scale parameters.
Now it is evident from the TB equation (\ref{TB}) that the scale parameters, which depend on plasma parameters such as $a$, $b$, $G_{e,p}$, and $N_{p}$, comprise all the information about the relaxed state. So the discriminant ($\Delta
=k_{2}^{2}k_{3}^{2}-4k_{2}^{3}+18k_{1}k_{2}k_{3}-4k_{1}k_{3}^{3}-27k_{1}^{2}$
) of equation (\ref{e13}) can be used to investigate the impact of
plasma parameters on the nature of scale parameters. When $\Delta <0,$ one of the
scale parameter is real and the other two are a complex conjugate pair; on the other hand, all the scale parameter are
real when $\Delta >0$. However, when $\Delta =0$, all the scale parameters
are real and at least two are equal.

Figure \ref{fig:1} illustrates the impact of plasma parameters such as Beltrami parameters, relativistic temperatures of pair species, and static ion species density on the nature of scale parameters by plotting $\Delta >0$ as a function of Beltrami parameters and relativistic temperature for various plasma settings. In the colored region, all the scale parameters are real and distinct, while in the transparent region, one scale parameter is real and the other two are complex conjugate pairs. For instance, the nature of scale parameters in a plasma setting with hot electrons, cold positrons ($G_{p}=1.2$) and a small fraction of static positive ion species ($N_{p}=0.9$ or $N_{i}=0.1$) is depicted in figure \ref{fig:1}(i). The analysis of the plot shows a clear relation between higher relativistic temperatures of electron species and a higher number of real-scale parameters. On the other hand, in figure \ref{fig:1}(ii), the analysis focuses on pure EP plasma ($N_{p}=1.0$), in which positrons are cold ($G_{p}=1.2$). In this particular instance, it is evident from the plot that, in the absence of immobile positive ion species, some real-scale parameters have transformed into complex ones. In figure \ref{fig:1}(iii), the investigation focuses on an EPI plasma, which has cold electrons ($G_{e}=1.2$) and a relatively low density of heavy ion species ($N_{p}=0.9$). Conversely, figure \ref{fig:1}(iv) examines an EP plasma ($N_{p}=1.0$) with cold electrons ($G_{e}=1.2$). Figures \ref{fig:1}(iii-iv) show that, even in the presence of cold electron species with hot positrons, the density of ion species significantly affects the nature of scale parameters, and complex scale parameters turn into real ones in the presence of small fraction of ion species in EP plasmas. An additional noteworthy observation that can be derived from the analysis of the plots presented in figures \ref{fig:1} (i-ii) and (iii-iv) pertains to the behavior of scale parameters in different scenarios. Specifically, it can be observed that in the scenario involving hot electrons with cold positrons, the scale parameters exhibit real values for lower values of $a$ in comparison to $b$. Conversely, in the scenario involving hot positrons with electrons, the scale parameters exhibit real values for lower values of parameter $b$ in comparison to parameter $a$. In the subsequent two cases, we examine a situation wherein the electron and positron species have the same temperatures, both in the presence or absence of static ions. Based on the data presented in figures \ref{fig:1}(v-vi), it is evident that the spectrum of real scale parameters exhibits a substantial enhancement. Additionally, the influence of ion density remains significant, even when temperatures are equal. So based on an analysis of figures \ref{fig:1}(i-vi) and the preceding discussion, it can be inferred that the inclusion of a small number of stationary ion species and the asymmetric temperatures of pair species have a substantial impact on the characteristics of the scale parameters.

The significance of the aforementioned analysis regarding the characteristics of scale parameters lies in the observation that relaxed states associated with real scale parameters typically exhibit a paramagnetic trend, while the presence of a combination of real and complex scale parameters shows a diamagnetic or partially diamagnetic trend \cite{Mahajan1998}.  In the relaxed state, diamagnetic structures have magnetic fields that increase from the center, while paramagnetic structures have magnetic fields that peak at or near the center and decrease towards the outer regions of the system. Therefore, the preceding analysis of the eigenvalues based on the discriminant of the cubic equation  (\ref{e13}) demonstrates the potential for the formation of both paramagnetic and diamagnetic structures in the plasma system, which can be controlled by temperature and density asymmetry in addition to generalized helicities. Moreover, it is also evident from figure \ref{fig:1} that when the Beltrami parameters of either electron or positron species are very small or one can say when flow of either species is super-Alfv\'{e}nic ($a$ or $b<1$), only one of the scale parameters is real while the other two are complex. In the singular limit, when the generalized helicities of both species vanish ($a$ and $b=0$) and flow vortices are aligned with the magnetic field then from equation (\ref{e13}), we get one scale parameter equal to zero while the other two are pure imaginary. This relaxed state with pair of imaginary eigenvalues will exhibit perfect diamagnetism \cite{Mahajan2008}.

\section{Analytical solution and field profiles}\label{S4}

As stated in the introduction, relativistic hot EPI plasmas are prevalent in a wide range of astrophysical environments, and there is a great deal of literature to support their existence. In addition, due to recent advancements in laser technology, trapping, and plasma confinement techniques, such plasmas can also be created in laboratories (for details, see \cite{Sarri2015,Stoneking2020,Fajans2020,Linden2021,Chen2023} and references therein).
For the numerical analysis of the relaxed state, we consider some arbitrary values of Beltrami parameters and relativistic temperatures while pair species density and ambient magnetic field typical to RT-1, in which plasma density is $10^{18}$cm$^{-3}$ and $B_{0}=10^{4}$gauss; corresponding to these values, skin depth and Alfv\'{e}n
velocity are $5.31\times10^{-4}$cm and  $9\times10^{7}$cm/sec, respectively \cite{Sato2023}. 
Now the general solution of the TB magnetic field equation (\ref{TB}) in an axisymmetric cylindrical geometry with an internal conductor can be written as
\begin{equation}
\mathbf{B}=\sum\limits_{j=1}^{3}\left( 
\begin{array}{c}
0 \\ 
C_{j}J_{1}(\lambda _{j}r)+D_{j}Y_{1}(\lambda _{j}r) \\ 
C_{j}J_{0}(\lambda _{j}r)+D_{j}Y_{0}(\lambda _{j}r)
\end{array}
\right),\label{AS}
\end{equation}
where $C_{j}$ and $D_{j}$ are arbitrary constants while $J_{\nu }$ and $%
Y_{\nu }$ are the Bessel functions of first and second kind and of order
zero and one \cite{Iqbal2001}. The values of $C_{j}$ and $D_{j}$ can be
computed with the help of following boundary conditions
\begin{equation}
\left\vert \mathbf{B}_{\theta }\right\vert _{r=d}=\sum\limits_{j=1}^{3}\left[
C_{j}J_{1}(\mu _{j})+D_{j}Y_{1}(\mu _{j})\right] =b_{1},  \label{c1}
\end{equation}
\begin{equation}
\left\vert \mathbf{B}_{z}\right\vert _{r=d}=\sum\limits_{j=1}^{3}\left[
C_{j}J_{0}(\mu _{j})+D_{j}Y_{0}(\mu _{j})\right] =b_{2},  \label{c2}
\end{equation}
\begin{equation}
\left\vert \left( \mathbf{\nabla }\times \mathbf{B}\right) _{z}\right\vert
_{r=d}=\sum\limits_{j=1}^{3}\lambda _{j}\left[ C_{j}J_{1}(\mu
_{j})+D_{j}Y_{1}(\mu _{j})\right] =b_{3},  \label{c3}
\end{equation}
\begin{equation}
\left\vert \left( \mathbf{\nabla }\times \mathbf{B}\right) _{\theta
}\right\vert _{r=d}=\sum\limits_{j=1}^{3}\lambda _{j}\left[ C_{j}J_{0}(\mu
_{j})+D_{j}Y_{0}(\mu _{j})\right] =b_{4},  \label{c4}
\end{equation}
\begin{equation}
\left\vert \lbrack \left( \mathbf{\nabla }\times \right) ^{2}\mathbf{B]}%
_{\theta }\right\vert _{r=d}=\sum\limits_{j=1}^{3}\lambda _{j}^{2}\left[
C_{j}J_{1}(\mu _{j})+D_{j}Y_{1}(\mu _{j})\right] =b_{5},  \label{c5}
\end{equation}
\begin{equation}
\left\vert \lbrack \left( \mathbf{\nabla }\times \right) ^{2}\mathbf{B]}
_{z}\right\vert _{r=d}=\sum\limits_{j=1}^{3}\lambda _{j}^{2}\left[
C_{j}J_{0}(\mu _{j})+D_{j}Y_{0}(\mu _{j})\right] =b_{6},  \label{c6}
\end{equation}
where $b_{j}$'s are arbitrary and real valued constants and $\mu _{j}=d\lambda
_{j}$. By solving the above equations (\ref{c1}-\ref{c6}) simultaneously, the following
values of constants $C_{j}$ and $D_{j}$ are obtained 
\begin{equation*}
C_{1}=\frac{\pi \mu _{1}\left[ Y_{1}\left( \mu _{1}\right) \left( \xi
_{1}-b_{6}\right) +Y_{0}\left( \mu _{1}\right) \left( \xi _{2}+b_{5}\right) 
\right] }{2\left( \lambda _{1}-\lambda _{2}\right) \left( \lambda
_{1}-\lambda _{3}\right) },
\end{equation*}
\begin{equation*}
C_{2}=\frac{\pi \mu _{2}\left[ Y_{0}\left( \mu _{2}\right) \left( \xi
_{3}-b_{5}\right) +Y_{1}\left( \mu _{2}\right) \left( \xi _{4}+b_{6}\right) 
\right] }{2\left( \lambda _{1}-\lambda _{2}\right) \left( \lambda
_{2}-\lambda _{3}\right) },
\end{equation*}
\begin{equation*}
C_{3}=\frac{\pi \mu _{3}\left[ Y_{0}\left( \mu _{3}\right) \left( \xi
_{5}-b_{5}\right) +Y_{1}\left( \mu _{3}\right) \left( \xi _{6}+b_{6}\right) 
\right] }{2\left( \lambda _{1}-\lambda _{3}\right) \left( \lambda
_{3}-\lambda _{2}\right) },
\end{equation*}
\begin{equation*}
D_{1}=\frac{\pi \mu _{1}\left[ J_{1}\left( \mu _{1}\right) \left( b_{6}-\xi
_{1}\right) -J_{0}\left( \mu _{1}\right) \left( \xi _{2}+b_{5}\right) \right]
}{2\left( \lambda _{1}-\lambda _{2}\right) \left( \lambda _{1}-\lambda
_{3}\right) },
\end{equation*}
\begin{equation*}
D_{2}=\frac{\pi \mu _{2}\left[ J_{0}\left( \mu _{2}\right) \left( b_{5}-\xi
_{3}\right) -J_{1}\left( \mu _{2}\right) \left( \xi _{4}+b_{6}\right) \right]
}{2\left( \lambda _{1}-\lambda _{2}\right) \left( \lambda _{2}-\lambda
_{3}\right) },
\end{equation*}
\begin{equation*}
D_{3}=\frac{\pi \mu _{3}\left[ J_{0}\left( \mu _{3}\right) \left( b_{5}-\xi
_{5}\right) -J_{1}\left( \mu _{3}\right) \left( \xi _{6}+b_{6}\right) \right]
}{2\left( \lambda _{1}-\lambda _{3}\right) \left( \lambda _{3}-\lambda
_{2}\right) },
\end{equation*}
where $\xi _{1}=\eta _{1}-\eta _{2}$, $\xi _{2}=\eta _{3}-\eta _{4}$, $\xi
_{3}=\eta _{5}-\eta _{6}$, $\xi _{4}=\eta _{7}-\eta _{8}$, $\xi _{5}=\eta
_{9}-\eta _{10}$, $\xi _{6}=\eta _{11}-\eta _{12}$, $\eta _{1}=\chi _{2}b_{4}
$, $\eta _{2}=\zeta _{2}b_{2}$, $\eta _{3}=\zeta _{2}b_{1}$, $\eta _{4}=\chi
_{2}b_{3}$, $\eta _{5}=\chi _{3}b_{3}$, $\eta _{6}=\zeta _{3}b_{1}$, $\eta
_{7}=\zeta _{3}b_{2}$, $\eta _{8}=\chi _{3}b_{4}$, $\eta _{9}=\chi _{1}b_{3}$
, $\eta _{10}=\zeta _{1}b_{1}$, $\eta _{11}=\zeta _{1}b_{2}$, $\eta
_{12}=\chi _{1}b$, $\chi _{1}=\lambda _{1}+\lambda _{2}$, $\chi _{2}=\lambda
_{2}+\lambda _{3}$, $\chi _{3}=\lambda _{1}+\lambda _{3}$, $\zeta
_{1}=\lambda _{1}\lambda _{2}$, $\zeta _{2}=\lambda _{2}\lambda _{3}$ and $%
\zeta _{3}=\lambda _{1}\lambda _{3}$. The values of boundary conditions and the radius of the
internal conductor to be used in the numerical analysis are $b_{1}=2.0$, $
b_{2}=1.6$, $b_{3}=1.0$, $b_{4}=0.5$, $b_{5}=0.4$, $b_{6}=0.3$ and $d=1$.
\begin{figure}
    \centering
    \includegraphics[scale=0.8]{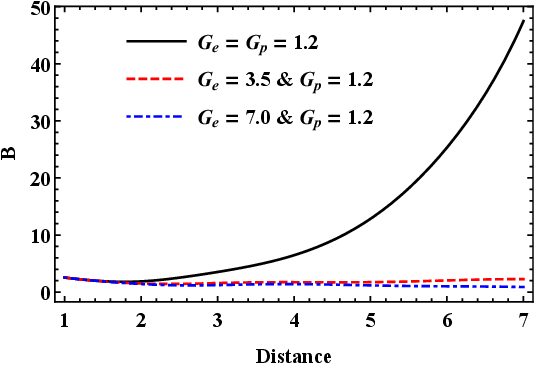}
    \caption{Radial profiles of the magnetic field for cold, slightly relativistic, and highly relativistic electron species ($G_{e}=1.2$, $3.5$ and $7.0$) when $a=1.5$, $b=3.0$, $G_{p}=1.2$ and $N_{p}=0.9$.}
    \label{fig:2}
\end{figure}

As the scale parameters of the TB state are
dependent on the five plasma parameters, namely relativistic temperatures ($G_{e,p}$), positron
density ($N_{p}$) and Beltrami parameters ($a,b$), we will now explore the impact of $G_{e,p}$ and $N_{p}$ on the\ self-organized magnetic field structures for the fixed values
of $a$ and $b$. These magnetic field structures have the potential to display either a paramagnetic or diamagnetic behavior. But here it is noteworthy to point out that the transition between paramagnetic structures and diamagnetic structures, and vice versa, depends on the nature of scale parameters. Particularly, when all three scale parameters are real, the analytical solution (\ref{AS}) involves a combination of Bessel functions of the first and second kind ($J_{\nu }$ and $Y_{\nu }$), which results in a decay away from the center and consequently shows a paramagnetic trend. On the other hand, when complex scale parameters are present along with a real one, $J_{\nu }$ and $Y_{\nu }$ are transformed into modified Bessel functions of the first and second kind ($I_{\nu }$ and $K_{\nu }$), respectively. In this case, a combination of $I_{\nu }$ and $K_{\nu }$ (for complex scale parameters), along with $J_{\nu }$ and $Y_{\nu }$ (for real scale parameter), leads to an increase away from the center and as a result plasma shows a diamagnetic trend. Furthermore, there are two significant implications arising from the transitions between paramagnetic and diamagnetic structures within the framework of our plasma system. One of the primary processes involves the conversion of magnetic energy to kinetic energy and vice versa. In the context of this plasma model, it is important to note that dissipative effects are not taken into account. Consequently, according to equation (\ref{energy}), it can be inferred that the conversion of magnetic energy into kinetic energy and vice versa is implied. The conversion of magnetic energy into kinetic energy bears resemblance to magnetic reconnection, a significant phenomenon observed in both astrophysical and laboratory plasmas. The second implication suggests that the diamagnetic state is advantageous for the confinement of plasma, whereas paramagnetic structures may lead to plasma disruption or deconfinement due to their relatively weaker values of magnetic field at the boundary of the plasma system.

\begin{figure}
    \centering
    \includegraphics[scale=0.8]{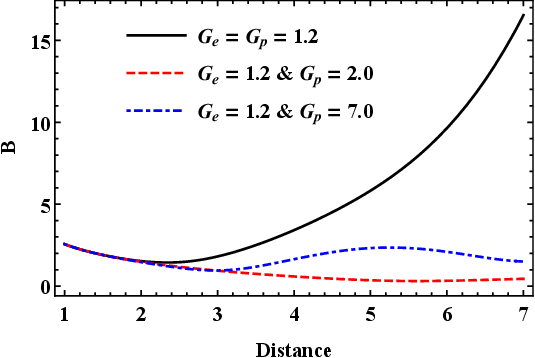}
    \caption{Radial profiles of the magnetic field for cold, slightly relativistic, and highly relativistic positron species ($G_{p}=1.2$, $2.0$ and $7.0$) when $a=-1.8$, $b=1.0$, $G_{e}=1.2$ and $N_{p}=0.9$.}
    \label{fig:3}
\end{figure}

\begin{figure}
    \centering
    \includegraphics[scale=0.8]{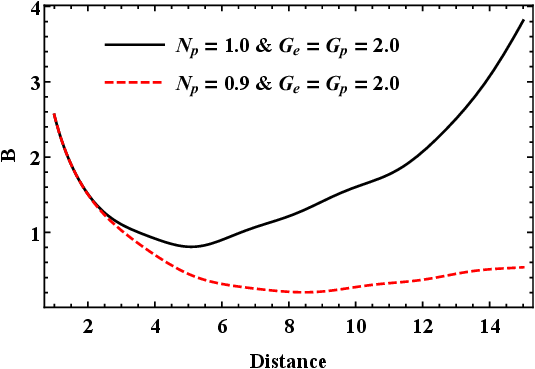}
    \caption{Radial profiles of the magnetic field for $N_{p}=1.0$ and $0.9$ when $a=-1.961$, $b=1.145$ and $G_{e}=G_{p}=2.0$.}
    \label{fig:4}
\end{figure}

Figures (\ref{fig:2}-\ref{fig:4}) shows a glimpse of impact of electron and
positron relativistic temperatures, and positron density or static ion species density for the given values of plasma
parameters on the relaxed magnetic field structures. So first we consider a set of plasma parameters: $a=1.5$, $b=3.0$ and $N_{p}=0.9$, with cold positrons $
G_{p}=1.2$ and
vary relativistic temperature of electron species ($G_{e}=1.2,3.5,7.0$). For these plasma parameters, when electrons are also cold i.e., $G_{e}=1.2$ and both pair species have the same temperatures ($G_{e}=G_{p}=1.2$), then the eigenvalues are $\lambda _{1}=2.7815$
and $\lambda _{2,3}=0.8592\pm 0.7516i$. In contrast, when the relativistic temperature of electron species is higher than that of positron species and its value is $G_{e}=3.5$ (slightly relativistic), all the scale parameters are real and there values are $\lambda _{1}=2.7480$, $\lambda _{2}=1.0905$ and $\lambda
_{3}=0.6614$. Similarly, when electrons are highly relativistic ($G_{e}=7.0$) and positrons are relatively cold ($G_{p}=1.2$), the scale parameters are again real and have the following values: $\lambda _{1}=2.7371$, $\lambda _{2}=1.3390$ and $\lambda
_{3}=0.4238$. Also note that the value of $\lambda _{1}$ almost remains the sames while the values of the $\lambda _{2}$ and $\lambda _{3}$ are changed. Figure (\ref{fig:2}) depicts the radial profiles of the magnetic field corresponding to the aforementioned plasma parameters and for different values of $G_{e}$. It is evident from the plot that when both species have lower and equal relativistic temperatures ($G_{e}=G_{p}=1.2$), the relaxed state exhibits diamagnetic behavior. On the contrary, when the relativistic temperature of electron species is higher than that of positron species (i.e., $G_{e}=3.5,7.0$), the field profiles show a paramagnetic trend. Additionally, the strength of the magnetic field has decreased when comparing $G_{e}=7.0$ to $G_{e}=3.5$. The observed trend can be attributed to an increase in the scale separation between the smallest (1/$\lambda _{1}$) and largest (1/$\lambda _{3}$) self-organized structures. Hence, it is evident from the preceding analysis and figure (\ref{fig:2}) that when positrons are cold and for certain values of Beltrami parameters, an increase in the temperature of electron species can transform the character of the relaxed state.

Next, we show the impact of the relativistic temperature of the positron species on the magnetic field structures by considering $a=-1.8$, $b=1.0$, $N_{p}=0.9$ and cold electrons  ($
G_{e}=1.2$) and
by changing the thermal energy of the positron species ($G_{p}=1.2,2.0,7.0$). Corresponding to these values of Beltrami parameters, electron species temperature, and positrons density, when both the electron and positron species are cold and have the same $G_{e}=G_{p}=1.2$, then one of the eigenvalues is real ($\lambda _{1}=-1.2828$) while the other two are conjugate pairs ($\lambda _{2,3}=0.2414\pm 0.5869i$). From figure (\ref{fig:3}), it can be seen that when both species have the same thermal energy as well as cold, there is a diamagnetic trend in the relaxed state.
Contrary to the above situation, 
when the relativistic temperature of positron species is greater than that of electron species and its value is $G_{p}=2.0$, all of the scale parameters are real, and their values are $\lambda _{1}=-1.2108$, $\lambda _{2}=-0.0425$ and $\lambda
_{3}=0.4533$. Corresponding to these real scale parameters, a paramagnetic trend of magnetic can be observed in figure (\ref{fig:3}). Similarly, when positrons are highly relativistic ($G_{p}=7.0$) and electrons are relatively cold ($G_{e}=1.2$), the scale parameters are once again real and have the following values: $\lambda _{1}=-1.0531$, $\lambda _{2}=-0.6399$ and $\lambda
_{3}=0.8931$. For these scale parameters, from figure (\ref{fig:3}), a paramagnetic trend can be seen for $G_{p}=7.0$. Also, the slight increase in the strength of the magnetic field for $G_{p}=7.0$ as compared to $G_{p}=2.0$ is due to the small scale separation between the scale parameters $\lambda _{1}$ and $\lambda _{3}$. Therefore, based on the aforementioned analysis of the figure (\ref{fig:3}), it is evident that for specific plasma parameters, an increase in temperature of the positron species can change the characteristics of the equilibrium state, particularly when the electrons are cold.

As mentioned in introduction that a relativistic hot EP plasmas can have a minority of ion species \cite{Berezhiani1994}. So, here we show that even a small density of immobile positive ions also have a significant impact on TB structures.
In this regard, figure (\ref{fig:4}) illustrates the impact of ion species density on the magnetic
structures for the fixed values of Beltrami parameters and thermal energies
($a=-1.961$, $b=1.145$ and $G_{e}=G_{p}=2.0$). Now, in the absence of ion species i.e.,
for pure EP plasma ($N_{p}=1.0$ or $N_{i}=0$), the eigenvalues are $\lambda _{1}=-1.6935$ and $\lambda
_{2,3}=0.4387\pm 0.2199i$. Since the scale parameters are combination of real and complex
values for $N_{p}=1.0$, so the self-organized magnetic structure shows
diamagnetic behavior.
However, when $N_{p}=0.9$ or $N_{i}=0.1$, the scale parameters are real and given by $\lambda
_{1}=-1.6906$, $\lambda _{2}=0.3483$ and $\lambda _{3}=0.5262$. As all of the
scale parameters are real, figure (\ref{fig:4}) depicts the formation of a
paramagnetic structure when $N_{p}=0.9$. Hence, from figure (\ref{fig:4}) and related data, one can conclude that even when pair species have equal temperatures, asymmetry in species density can transform a diamagnetic structure into a paramagnetic one and vice versa.

The preceding discussion leads one to the conclusion that for specific values of Beltrami parameters when the pair species are cold and have equal temperatures, one of the scale parameters is real while the other two are complex, and consequently, the TB relaxed state shows a diamagnetic trend. In contrast, when there is temperature asymmetry between pair species, such that when either plasma species has a higher relativistic temperature, all three scale parameters are real and self-organized state exhibits paramagnetism. In addition, the analysis also demonstrates that even when both plasma species have the same relativistic temperature, a small fraction of ion species can transform the diamagnetic trend into a paramagnetic one. So it can be concluded that for certain values of generalized helicities of pair species, asymmetry in temperature and density of pair species can transform the nature of the relaxed state structures. So the formulation and the results presented in this study have the potential to provide valuable insights into the equilibrium structures observed in astrophysical objects, specifically the pulsar magnetospheres \cite{Petri2016}. Additionally, it will have implications for future laboratory experiments involving the utilization of internal conductor plasma configurations such as those in RT-1, galatea traps, and MRX \cite{Yoshida2006,Yoshida2010b,Sato2023,Yamada1997,Morozov2007}.
\section{Summary}\label{S5}
The relaxed state of a relativistic two-temperature electron-positron plasma with a small fraction of static positively charged ions has been investigated. Starting from relativistic equations of motion for electron and positron species, the generalized vortex evolution equations are obtained. The steady-state solution of the generalized vortex evolution equations yields two Beltrami conditions. By manipulating these Beltrami conditions along with Ampere's law, a TB state equation is derived. The TB state is a non-force-free state that shows strong magnetofluid coupling. Also, the TB relaxed state is a linear combination of three force-free fields with three scale parameters, resulting in the formation of three self-organized structures. The analysis of scale parameters reveals that their characteristics can be changed as a result of temperature and density asymmetry. In the case of lower relativistic temperatures and ion densities, as well as for smaller values of Beltrami parameters, it has been observed that one of the scale parameters is real while the other two form a complex conjugate pair. Importantly, the increase in thermal energies of either electrons or positrons and the small number of stationary ions in EP plasma tend to change some of the complex scale parameters into real ones. Moreover, the analytical solution of the TB state in an axisymmetric cylindrical geometry with an internal conductor at its center is also presented.
It is investigated that for some specific values of Beltrami parameters and ion density when both plasma species are cold, one scale parameter is real while the other two are complex conjugate pairs, and corresponding to these scale parameters, radial profiles of the magnetic field show a diamagnetic trend. The increase in thermal energies of either the electron or positron species tends to convert complex scale parameters into real ones. As a result of this transition in scale parameters, diamagnetic structures become paramagnetic. Importantly, the study reveals that self-organized structures are paramagnetic when one of the species has a higher relativistic temperature than the other. Furthermore, it is also observed that in paramagnetic structures, the magnitude of the magnetic field experiences a small change as a result of the higher relativistic temperature of plasma species. The study also demonstrates the significance of density asymmetry in electron-positron (EP) plasma by examining the behavior of the TB state for certain fixed values of Beltrami parameters and equal temperatures of pair species. Specifically, it is observed that pure EP plasma exhibits diamagnetism in the TB state. However, the presence of a small fraction of ion species in EP plasma (referred to as EPI) leads to a transition from diamagnetism to paramagnetism. Therefore, the presence of even a minimal concentration of static positive ions exerts a significant impact on the relaxed state. The significance of paramagnetic and diamagnetic structures, as well as the transitions between them resulting from temperature and density asymmetries, holds relevance within the realm of plasma confinement and energy transformation mechanisms. Therefore, the findings presented here will be helpful in attaining a better understanding of relativistic hot pair plasmas in laboratory experiments that involve internal conductor configurations as well as equilibrium structures in cosmic environments such as pulsar magnetospheres.
\subsection*{Research funding}
The work of M. Iqbal is funded by Higher Education Commission (HEC),
Pakistan under project No. 20-9408/Punjab/NRPU/R\&D/HEC/2017-18.
\subsection*{Author Contributions}
Both authors have accepted responsibility for the entire content of this submitted manuscript and approved submission.
\subsection*{Conflict of interest statement}
The authors declare no conflicts of interest regarding this article.
\subsection*{Data Availability Statement}
Data sharing is not applicable to this article as no datasets were generated or analyzed during the current study.
\bibliography{biblo}

\end{document}